\documentclass[aps]{revtex4}

\newcommand{\BB}[1]		{\mbox{\boldmath${#1}$}}

\newcommand{\Ve}		{{\BB{e}}}
\newcommand{\Vf}		{{\BB{f}}}
\newcommand{\Vg}		{{\BB{g}}}

\newcommand{\Vq}		{{\BB{q}}}

\newcommand{\Vv}		{{\BB{v}}}

\newcommand{\VH}		{{\BB{H}}}

\newcommand{\DX}		{\BB{\nabla}}

\newcommand{\Dx}		{\partial_x}
\newcommand{\Dz}		{\partial_z}
\newcommand{\Dt}		{\partial_t}

\newcommand{\mm}		{\, \rm mm}
\newcommand{\cm}		{\, \rm cm}

\newcommand{\s}			{\, \rm s}

\newcommand{\cf}		{{\it c.f. }}
\newcommand{\ie}		{{\it i.e. }}

\usepackage{graphicx}
\usepackage{natbib}
\usepackage{amssymb,amsmath,empheq}

\newcommand{\Da}			{{\rm Da}}
\newcommand{\Pe}			{{\rm Pe}}
\newcommand{\Sh}			{{\rm Sh}}
\newcommand{\Hc}			{{\hat c}}
\newcommand{\Hh}			{{\hat h}}
\newcommand{\Hq}			{{\hat q}}
\newcommand{\Hu}			{{\hat u}}
\newcommand{\Hv}			{{\hat v}}
\newcommand{\Ho}			{{\hat \omega}}
\newcommand{\tG}			{{\tilde G}}
\newcommand{\tH}			{{\tilde H}}

\begin{document}

\author{Piotr Szymczak$^1$ and Anthony J. C. Ladd$^2$}
\affiliation{$^1$Institute of Theoretical Physics, Faculty of Physics, University of Warsaw, Ho\.{z}a 69, 00-618, Warsaw, Poland\\
$^2$Chemical Engineering Department, University of Florida, Gainesville, FL  32611-6005, USA}

\title{Reactive-infiltration instabilities in rocks. Fracture dissolution}

\begin{abstract}
A reactive fluid dissolving the surface of a uniform fracture will trigger an instability in the dissolution front, leading to spontaneous formation of pronounced well-spaced channels in the surrounding rock matrix. Although the underlying mechanism is similar to the wormhole instability in porous rocks there are significant differences in the physics, due to the absence of a steadily propagating reaction front. In previous work we have described the geophysical implications of this instability in regard to the formation of long conduits in soluble rocks. Here we describe a more general linear stability analysis, including axial diffusion, transport limited dissolution, non-linear kinetics, and a finite length system.
\end{abstract}

\maketitle

\section{Introduction}

Fracture dissolution is an important component of a number of geological processes, including the early stages of karstification \citep{Hanna1998}, diagenesis \citep{Laubach2010}, and the evolution of carbonate aquifers \citep{Ortoleva1994}.  It also plays an important role in geoengineering applications such as dam stability \citep{Romanov2003}, oil reservoir stimulation methods~\citep{Economides2000} and  leakage of sequestered ${\rm CO}_2$~\citep{Pruess2008}. The dynamics of evolving fractures is complex, due to the highly nonlinear couplings between morphology, flow and dissolution. Theoretical~\citep{Hanna1998,Detwiler2007,Szymczak2011} and experimental studies~\citep{Durham2001,Gouze2003,Detwiler2003} have shown that the positive feedback between fluid transport and mineral dissolution leads to an instability in an initially uniform reaction front and the subsequent formation of pronounced dissolution channels, deeply etched into the rock surfaces. These processes were shown to be important in the development of limestone caves~\citep{Szymczak2011}, and also in the assessment of subsidence hazards, since they dramatically speed up the growth of long conduits. Understanding spontaneous flow focusing during fracture dissolution is also important to the petroleum industry, for efficient acidization of natural fractures and for acid fracturing of porous rocks. In the former process, acid is pumped into the fractured reservoir to dissolve material  blocking the pathways between the wellbore and the reservoir. Spontaneous channeling increases the effectiveness of the process by creating highly permeable pathways, minimizing the amount of acid needed. In acid fracturing the fluid pressure is high enough to induce hydrofracturing; the newly created fractures are then etched with acid to increase the permeability of the system. Nonuniform dissolution is crucial in this process, since a uniformly etched fracture will close tightly under the overburden once the fluid pressure is removed; significant permeability will only be created by inhomogeneous etching when the less dissolved regions act as supports to keep more dissolved regions open.

In this paper we investigate the initiation of the instability in a fracture dissolution front and assess the wavelength and growth rate of the most unstable mode as a function of physical parameters characterizing the rates of transport and reaction in the fracture. In Sec.~\ref{sec:eqn} we present the two-dimensional averaged equations for fracture dissolution; a detailed justification of the transport equation \eqref{eq:CD} is given in Appendix~\ref{sec:CD}. Next we consider a uniform fracture where an analytic solution is possible; this forms the base state for the subsequent stability analysis in Sec.~\ref{sec:lsa}. Results are presented in Sec.~\ref{sec:results}, extending our previous analysis~\citep{Szymczak2011} in several directions. We now consider axial diffusion of reactant as well as lateral diffusion and also the effect of cross-aperture diffusion on the effective reaction rate. After that we lift the assumptions that the fracture is of infinite length and that the reaction kinetics are linear. We finish with a summary of our results and conclusions. In a subsequent paper we will describe an  analysis of the instability in the dissolution of a porous matrix.

\section{Equations for fracture dissolution}\label{sec:eqn}

\begin{figure}
\center\includegraphics[width=0.8\textwidth]{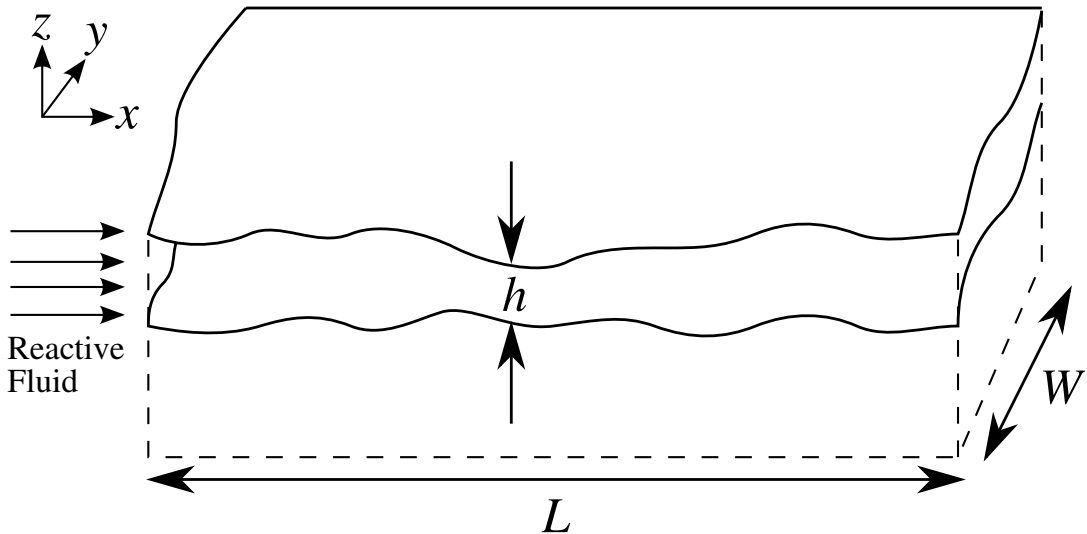}
\caption{Dissolution of a rough fracture of length $L$ and width $W$; fluid flow is in the $x$ direction and the fracture surfaces dissolve in the normal ($z$) direction. The aperture $h(x,y)$ is the distance between the fracture surfaces.}\label{fig:fig1}
\end{figure}

Fractures are geometrically characterized by a short dimension ($z$ direction), the aperture, and two much longer dimensions, length ($x$ direction) and width ($y$ direction). In natural fractures the aperture is typically less than $1 \mm$, while the length ($L$) and width ($W$) are of the order of meters (see Fig.~\ref{fig:fig1}). It is typical to exploit this difference in scales by
introducing approximate two-dimensional equations for fluid flow, reactant transport, and erosion. Fluid flow is described by the Reynolds equation for the local volume flux (per unit length across the fracture), $\Vq(x,y,t) = \int_0^h \Vv(x,y,z,t) dz$:
\begin{equation}\label{eq:Reynolds}
\Vq = - \frac{h^3}{12 \mu} \DX p, ~~ \DX \cdot \Vq = 0,
\end{equation}
where $\mu$ is the fluid viscosity. The essence of the Reynolds approximation is to assume that the exact result for stationary flow between parallel plates can be applied locally to a varying aperture. In this approximation the pressure is independent of height and reduces to the two-dimensional field $p(x,y)$. The validity of the Reynolds approximation for rough fractures has been examined in \cite{Oron1998} and \cite{Berkowitz2002}. The key requirements are: (i) low Reynolds number flow, $Re \ll 1$ (ii) slow variation in aperture $\left | \nabla h \right | \ll 1$. We will assume these conditions hold in what follows. The incompressibility condition in Eq.~\eqref{eq:Reynolds} ignores effects of the reactant (or product) concentration on the mass density of the fluid. This assumption is valid for the majority of natural systems; for example, in limestone dissolution the density correction due to the dissolved species is of the order of 0.01\%. However, dissolution of halite (rock salt) is a notable exception; here the increase in mass density can be as large as 25\%. 

The transport of reactant can be described in terms of a two-dimensional concentration field that has been averaged over the aperture. The most important average is the ``cup-mixing'' or velocity-averaged concentration~\citep{Bird2001},
\begin{equation}\label{eq:cupc}
c(x,y,t) = \frac{1}{\left|\Vq(x,y,t)\right|} \int_0^{h(x,y,t)} \left|\Vv(x,y,z,t)\right| c_{3d}(x,y,z,t) dz,
\end{equation}
where we use $c_{3d}$ to identify the three-dimensional concentration field. Under certain conditions, discussed in Appendix~\ref{sec:CD}, the three-dimensional convection-diffusion equation for reactant transport in the fracture can be reduced to a two-dimensional convection-diffusion-reaction equation for the cup-mixing concentration~\citep{Hanna1998,Detwiler2007,Szymczak2011},
\begin{equation}\label{eq:CD}
\Vq \cdot \DX c = D \DX h \cdot \DX c - 2R(c),
\end{equation}
where $R(c)$ accounts for reactant transfer at each of the fracture surfaces. The slow dissolution of the rock surfaces allows the time-dependence in Eq.~\eqref{eq:CD} to be neglected (Appendix~\ref{sec:SS}).

In this paper we will usually assume a first-order dissolution reaction at the fracture surfaces $R = kc_w$, where $k$ is the rate constant and $c_w$ is the reactant concentration at the fracture surface. The reactive flux $R$ must balance the diffusive flux at the surface
\begin{equation}
\label{eq:Rdiff}
R_{diff} = -D (\nabla c)_w,
\end{equation}
where the gradient is pointing towards the surface. Alternatively, and more usefully, the diffusive flux can be expressed in terms of the difference between the surface concentration, $c_w$, and the cup-mixing concentration, $c$ by using a mass-transfer coefficient or Sherwood number~\citep{Bird2001},
\begin{equation}
\label{eq:RdiffSh}
R_{diff} = \frac{D \Sh}{2 h}(c-c_w).
\end{equation}
The Sherwood number, $\Sh$, depends on reaction rate at the fracture surfaces ($k$) but the variation is relatively small~\citep{Hayes1994,Gupta2001}, bounded by two asymptotic limits: high reaction rates (transport limit), $\Sh=7.54$, and low reaction rates (reaction limit), $\Sh=8.24$. In the numerical calculations we approximate the Sherwood number by a constant value $\Sh=8$.

By equating the reactive and diffusive fluxes $R = R_{diff}$ we obtain the standard relationship between $c_w$ and $c$ \citep{Gupta2001},
\begin{equation}\label{eq:CW}
c_w = \dfrac{c}{1+2kh/D\Sh}.
\end{equation}
The reactive flux can then be expressed in terms of the cup-mixing concentration,
\begin{equation}\label{eq:Rcm}
R(c) = k_{eff}c,
\end{equation}
where the effective reaction rate is given by
\begin{equation}\label{eq:keff}
k_{eff}(h) = \dfrac{k}{1+2kh/D\Sh},
\end{equation}
In sufficiently narrow apertures the dissolution kinetics are reaction limited and the concentration field is almost uniform across the aperture so that $k_{eff} \approx k$. However, as the fracture opens the reaction rate becomes hindered by diffusive transport of reactant across the aperture. When $k h/D Sh \gg 1$, dissolution can become entirely diffusion limited with $k_{eff} \approx D\Sh/2h$.

A derivation of Eq.~\eqref{eq:CD}, with the kinetics described in Eqs.~\eqref{eq:Rcm} and \eqref{eq:keff}, will be given in Appendix~\ref{sec:CD}, starting from the full three-dimensional transport equations. In particular, the diffusive term in Eq.~\eqref{eq:CD} is shown to be purely molecular for either convective ($q/D \rightarrow \infty$) or reaction-limited ($2kh/D \rightarrow 0$) transport. In taking the Sherwood number to be independent of the distance from the inlet, we are assuming that entrance effects are negligible. For a flat plat geometry the entrance length scale $l_{in}$ is given by \citep{Ebadian1998}
\begin{equation}\label{eq:Lin}
l_{in} = 0.016\frac{q h}{D},
\end{equation}
taking $l_{in}$ as the distance over which the Sherwood number is within $5\%$ of its asymptotic value. This length is small compared to the reactant penetration length under the typical conditions of fracture dissolution (see Sec.~\ref{sec:1D}).

Equations \eqref{eq:Rcm} and \eqref{eq:keff} describe a dissolution reaction controlled by the concentration of reactant; a typical example is dissolution of fractures (or porous rocks) by a strong acid. However, when calcite is dissolved by aqueous ${\rm CO}_2$ at pH values similar to those of natural groundwater, the dissolution rate is limited by the calcium ion undersaturation $c_{sat} - c_{ca}$ \citep{Plummer1978},
\begin{equation}\label{eq:Rcunder}
R(c_{ca}) = -k_{eff}(c_{sat} - c_{ca}),
\end{equation}
where $c_{ca}$ is the flow-averaged concentration of dissolved calcium ions. The sign of $R$ accounts for a dissolution flux into the fluid rather than a reactive flux into the surface and so the transport equation for the undersaturation takes the same form as \eqref{eq:CD}. In the rest of the paper we will use $c$ to represent either the concentration of reactant or the undersaturation of dissolved minerals.

A reactive fluid with an inlet ($x = 0$) concentration $c_{in}$ dissolves the surrounding rock, increasing the fracture aperture at a rate
\begin{equation}
\label{eq:erosion}
\Dt h = 2 k_{eff} \gamma \frac{c}{c_{in}},
\end{equation}
where $\gamma = c_{in}/\nu c_{sol}$ is the acid capacity number or volume of solid dissolved by a unit volume of reactant. Here $c_{sol}$ is the molar concentration of soluble material and $\nu$ accounts for the stoichiometry of the reaction. Mineral concentrations in the solid phase, are typically much higher than reactant concentrations in the aqueous phase and the characteristic dissolution time,
\begin{equation}\label{eq:td}
t_d = h/2k_{eff}\gamma,
\end{equation}
is large for natural minerals in typical groundwater conditions; for limestone fractures it is approximately 2 months \citep{Szymczak2011}. Thus there is a significant separation between the dissolution time scale and the relaxation of the concentration field ($t \sim h^2/D$), which justifies dropping the time
dependence in Eq.~\eqref{eq:CD}; for further discussion see Appendix~\ref{sec:SS}.

\section{Concentration profile in a uniform fracture}\label{sec:1D}

Let us first consider a uniform aperture $h(x,y)=h_0$ and find the corresponding concentration profile; the solutions will form the base state for the stability analysis. The flow rate $q_0$ is independent of space and the transport equation is
\begin{equation}\label{eq:CD1D}
q_0 \partial_x c - D h_0 \partial_x^2 c = - \frac{2 kc}{1 + G},
\end{equation}
where we have absorbed the transport correction into a single factor,
\begin{equation}\label{eq:GG}
G=\frac{2 k h_0}{D\Sh}.
\end{equation}

For an inlet concentration $c_{in}$, Eq.~\eqref{eq:CD1D} has an exponentially decaying solution,
\begin{equation}\label{expo}
c(x) = c_{in}e^{-\kappa x},
\end{equation}
with a penetration length $l_p = \kappa^{-1}$ given by
\begin{equation}
\kappa h_0 = \frac{\Pe}{2} \left(\sqrt{1+\frac{4 \Da_{eff}}{\Pe}}-1\right).
\end{equation}
The P\'eclet number,
\begin{equation}
\Pe=\frac{q_0}{D},
\label{Pe1}
\end{equation} measures the relative magnitude of convective and diffusive transport of solute, and the effective Damk\"{o}hler number,
\begin{equation}
\Da_{eff}=\frac{2k_{eff} h_0}{q_0} = \frac{2k h_0}{(1+G)q_0},
\label{Da1}
\end{equation}
relates the effective surface reaction rate, Eqs.~\eqref{eq:Rcm} and \eqref{eq:keff}, to the rate of convective transport.

It will be convenient to frame our results in terms of the transport correction $G$ \eqref{eq:GG} and the convective parameter
\begin{equation}\label{eq:HH}
H=\frac{\Da_{eff}}{\Pe}.
\end{equation}
A discussion of the natural length scales of the problem and their relation to $H$ can be found in Appendix \ref{sec:A2}. 
The inverse penetration length can be written in terms of $H$,
\begin{equation}\label{eq:kappa}
\kappa h_0 = \frac{\Pe}{2} \left(\sqrt{1+4 H}-1\right),
\end{equation}
with the important limiting cases:
\begin{itemize}
\item convection dominated ($H \rightarrow 0$)
\begin{equation}\label{h0}
\kappa h_0 =\Da_{eff},
\end{equation}
\item diffusion dominated ($H \rightarrow \infty$)
\begin{equation}\label{hinf}
\kappa {h_0} = \sqrt{\Pe \Da_{eff}} = \sqrt{\frac{G \Sh}{1 + G}},
\end{equation}
\end{itemize}
In Appendix~\ref{sec:CD} we show that \eqref{eq:CD} is valid for all $G$ when $H = 0$ (Sec. \ref{sec:conv}) and for all $H$ when $G \ll 1$ (Sec. \ref{sec:rlim}). 

For long fractures, the reactant penetration length is the natural length scale for dissolution. On the scale of $\kappa^{-1}$ the entrance length \eqref{eq:Lin} is
\begin{equation}\label{eq:Lins}
\kappa l_{in} = 0.008\,\Pe^2(\sqrt{1+4H}-1).
\end{equation}
In the convective ($H \rightarrow 0$) limit, $\kappa l_{in} = 0.016G \Sh/(1+G) < 0.12$ over the whole range of reaction rates; it is vanishingly small in the reaction ($G \rightarrow 0$) limit. In the diffusive ($H \rightarrow \infty$) limit $\kappa l_{in} = 0.016 \Pe \sqrt{G \Sh/(1+G)} < 0.05 \Pe$, which is again small (since $\Pe \ll 1$). In Sec.~\ref{sec:L} we will examine the instability in finite-length fractures $\kappa L < 1$, but only in the reaction limit ($G \rightarrow 0$), in which case $l_{in}/L \rightarrow 0$, even for finite $\kappa L$.

\section{Linear stability analysis of a uniform profile}\label{sec:lsa}

The discussion in Sec.~\ref{sec:eqn}, supported by the derivations in Appendix~\ref{sec:CD}, leads to the following average equations for the concentration, aperture and flow fields in an evolving fracture:
\begin{align}
& q_x \partial_x c + q_y \partial_y c  - D\left[\partial_x (h \partial_x c) + \partial_y (h \partial_y c)\right] = - \frac{c_{in}}{\gamma} \partial_t h  & \text{(transport)} 
\label{eq:transport} \\
& c_{in}\partial_t h= \frac{2 k \gamma c}{1+ 2 kh/D\Sh}\ \ & \text{(erosion)} \label{eq:erosion2}\\
& \partial_x q_x + \partial_y q_y = 0 & \text{(continuity)} 
\label{eq:continuity} \\
& \partial_y q_x - \frac{3}{h} q_x \partial_y h  = \partial_x q_y - \frac{3}{h} q_y
\partial_x h  \ \ & \text{(compatibility)}
\label{eq:compatibility}
\end{align}
Here the Reynolds equation \eqref{eq:Reynolds} has been replaced by the more convenient equations for continuity \eqref{eq:continuity} and compatibility \eqref{eq:compatibility} (see Appendix~\ref{compder}). When supplemented by appropriate boundary conditions:
\begin{equation}\label{eq:cbc}
c(x=0,y,t)=c_{in}, \ \ \ \ c(x\rightarrow \infty,y,t)=0,
\end{equation}
\begin{equation}\label{eq:qbc}
q_x(x\rightarrow \infty,y,t)=q_0, \ \ \ \ q_y(x=0,y,t)=0,
\end{equation}
Eqs.~\eqref{eq:transport}--\eqref{eq:compatibility} form a complete, albeit approximate, description of the erosion of a single fracture (in the domain $x>0$).
The constant pressure condition at the inlet has been replaced by the boundary condition $q_y(x=0) = 0.$

The above equations allow one-dimensional solutions in which the fields depend only on $x$ and $t$. This corresponds to uniform dissolution of the fracture, an assumption still commonly found in models of fracture dissolution~\citep{Dreybrodt1990,Dreybrodt1996}. For example, in the reaction-limited, convection-dominated case ($G \rightarrow 0$, $H \rightarrow 0$), the solution is
\begin{equation}
c(x,t) = c_{in} e^{- 2k x/q},
\label{cbase}
\end{equation}
\begin{equation}
h(x,t) = h_0 + 2 k \gamma t e^{- 2k x/q},
\label{hbase}
\end{equation}
\begin{equation}
{\Vq}(x,t) = q_0 {\Ve}_x.
\label{qbase}
\end{equation}
In \cite{Szymczak2011} we showed that the solution represented by Eqs.~\eqref{cbase}--\eqref{qbase} is unstable to infinitesimal perturbations along the $y$ direction. Here we will not limit ourselves to the reaction-limited, convection dominated regime, but consider more general kinetics and transport. Thus $\kappa$ will no longer be equal to $2k/q$, as in \eqref{cbase} and \eqref{hbase}, but instead it will be given by the general expression \eqref{eq:kappa}.

An important detail in the stability analysis is that the base state for the aperture \eqref{hbase} is itself time-dependent. The stability of nonautonomous systems is in general a difficult problem \citep{Farrell1996} and in \cite{Szymczak2011} we adopted an approximate approach~\citep{Tan1986} in which the base state is frozen at a specific time, $t_0$, and the growth rate is then determined as if the base state were time-independent (the quasi-steady-state approximation). The validity of this approach was tested by comparing the results of the quasi-steady-state approximation with a numerical solution of the complete system of equations \eqref{eq:transport}--\eqref{eq:compatibility}. In particular, we were able to show that the most relevant instability is obtained by freezing the base state at $t_0=0$ and in the present paper we will focus on this case. 
The solution at $t=0$ is
\begin{equation}
c_b(x) = c_{in} e^{- \kappa x}, \ \ \ \ h_b(x)=h_0, \ \ \ \ {\Vq}_b(x) = q_0 {\Ve}_x,
\label{base}
\end{equation}
which simplifies the subsequent calculations. 

The linear stability analysis proceeds by considering infinitesimal perturbations to the base profile \eqref{base}: $h=h_b+\delta h$, $c=c_b + \delta c$ and ${\Vq} = {\Vq}_b + \delta {\Vq}$. This gives the following linearized equations for the aperture, concentration and flow fields:
\begin{equation}
\delta q_x \partial_x c_b + q_b \partial_x \delta c - D \left[h_b  \partial_x^2 \delta c + h_b \partial_y^2 \delta c + \delta h \partial_x^2 c_b + (\partial_x \delta h) (\partial_x c_b)\right] = - \frac{c_{in}}{\gamma} \partial_t \delta h,
\label{transport3}
\end{equation}
\begin{equation}
c_{in}\left(1+ \frac{2 k h_b}{D\Sh} \right)  \partial_t \delta h +  \left(1+ \frac{2 k h_b}{D\Sh} \right)^{-1} \frac{(2 k)^2  \gamma c_b}{D\Sh} \delta h = 2 k \gamma \delta c,
\label{erosion3}
\end{equation}
\begin{equation}
\partial_x \delta q_x + \partial_y \delta q_y = 0, 
\label{cont3}
\end{equation}
\begin{equation}
\partial_y \delta q_x - \frac{3}{h_b} q_b \partial_y \delta h  = \partial_x  \delta q_y.  
\label{compat3}
\end{equation}
Terms in $\partial_x h_b$ have been omitted from Eqs.~\eqref{transport3} and \eqref{compat3}, since the expansion is about an $x-$independent aperture field. In Eq.~\eqref{erosion3} we have made use of the erosion equation for the base field, $c_{in} (1+2kh_b/DSh) \partial_t h_b = 2 k \gamma c_b$.

The linearized equations for fracture dissolution can be simplified by transforming to dimensionless variables. We take the penetration length $\kappa^{-1}$ as the unit of length, and the characteristic inlet dissolution time, $t_d$ \eqref{eq:td}, as the unit of time. The dimensionless variables are then:
\begin{equation}\label{dims1a}
\xi = \kappa x, \ \ \ \ \ \eta = \kappa y, \ \ \ \ \ \tau = \frac{2k\gamma t}{(1+G)h_0}.
\end{equation}
The concentration is scaled by the inlet concentration $c_{in}$, while the aperture and flow rate are scaled by their (constant) values in the base state:
\begin{equation}\label{dims2}
\Hc=\frac{c}{c_{in}}, \ \ \ \ \Hh = \frac{h}{h_0}, \ \ \ \ {\hat \Vq}=\frac{\Vq}{q_0}.
\end{equation}
The dimensionless base-state solution is:
\begin{equation}
\Hc_b=e^{-\xi}, \ \ \ \ \Hh_b = 1, \ \ \ \ {\hat \Vq}_b = \Ve_\xi,
\end{equation}
and the dimensionless perturbations can be found from the following equations:
\begin{equation}
\frac{2k}{q_0\kappa(1+G)} \partial_\tau \delta \Hh = e^{-\xi} \delta \Hq_\xi - \partial_\xi \delta \Hc + \frac{D\kappa h_0}{q_0} \left(\partial_\xi^2 \delta \Hc + \partial_\eta^2 \delta \Hc +  e^{-\xi} \delta \Hh - e^{-\xi} \partial_\xi \delta \Hh \right),
\label{transport4}
\end{equation}
\begin{equation}
\partial_\tau \delta \Hh +  \frac{G}{1+G} e^{-\xi} \delta \Hh = \delta \Hc,
\label{erosion4}
\end{equation}
\begin{equation}
\partial_\xi^2 \delta \Hq_\xi + \partial_\eta^2 \delta \Hq_\xi = 3\partial_\eta^2 \delta \Hh.
\label{cont4}
\end{equation}
In deriving \eqref{cont4} we have combined the continuity equation \eqref{cont3} and the compatibility equation \eqref{compat3} to eliminate $\delta \Hq_\eta$.

The transport equation \eqref{transport4} involves two new dimensionless constants, each one based on the penetration length $\kappa^{-1}$,
\begin{eqnarray}
\Pe_\kappa &=& \dfrac{q_0}{D \kappa h_0} = \frac{2}{\sqrt{1+4H}-1}, \label{eq:Pek} \\
\Da_\kappa &=& \frac{2k_{eff}}{q_0 \kappa} = \frac{2H}{\sqrt{1+4H}-1}. \label{eq:Dak}
\end{eqnarray}
$\Pe_\kappa$ is the ratio of convective to diffusive fluxes on the length scale $\kappa^{-1}$, while $\Da_\kappa$ is the ratio of convective to reactive fluxes on the same scale. The physical significance of these parameters is discussed in Appendix \ref{sec:A2}. Rewriting the transport equation in terms of $\Pe_\kappa$ and $\Da_\kappa$ and rearranging to isolate the term in $\delta \Hq_\xi$,
\begin{equation}
\delta \Hq_\xi =  e^\xi \left[\Da_\kappa \partial_\tau + \Pe_\kappa^{-1}\partial_\xi e^{-\xi} \right]\delta \Hh + e^\xi \left[\partial_\xi - \Pe_\kappa^{-1} (\partial_\xi^2 + \partial_\eta^2) \right]\delta \Hc .
\label{transport4a}
\end{equation}

Assuming that the perturbations are sinusoidal in $\eta$ and exponential in $\tau$,
\begin{eqnarray}
\delta \Hc &=& f_c(\xi)\cos(\Hu\eta)e^{\Ho \tau}, \label{eq:dc} \\
\delta \Hh &=& f_h(\xi)\cos(\Hu\eta)e^{\Ho \tau}, \label{eq:dh} \\
\delta \Hq_\xi &=& f_q(\xi)\cos(\Hu\eta)e^{\Ho \tau}. \label{eq:dq}
\end{eqnarray}
Note that $\Ho$ and $\Hu$ are dimensionless quantities related to the instability growth rate $\omega$ and wavelength $\lambda$ by the relations
\begin{equation}
\Ho = \omega t_d, ~~ \Hu = \frac{2\pi}{\kappa\lambda}.\label{eq:hatdef}
\end{equation}

Substituting the expansions \eqref{eq:dc}-\eqref{eq:dq} into Eqs.~\eqref{transport4a}, \eqref{erosion4}, and \eqref{cont4} leads to coupled equations for the one-dimensional fields $f_c(\xi)$, $f_h(\xi)$, and $f_q(\xi)$:
\begin{equation}
f_q =  e^\xi \left[\Da_\kappa \Ho + \Pe_\kappa^{-1}\partial_\xi e^{-\xi} \right]f_h + e^\xi \left[\partial_\xi - \Pe_\kappa^{-1} (\partial_\xi^2 - \Hu^2) \right]f_c.
\label{transport5}
\end{equation}
\begin{equation}
\left(\Ho +  \frac{Ge^{-\xi}}{1+G}\right)f_h = f_c.
\label{erosion5}
\end{equation}
\begin{equation}
(\partial_\xi^2 - \Hu^2)f_q = -3\Hu^2f_h.
\label{cont5}
\end{equation}
Eliminating $f_c$, we express $f_q$ in terms of $f_h$ only
\begin{equation}
f_q = e^\xi \left\{\left[\Da_\kappa \Ho + \Pe_\kappa^{-1}\partial_\xi e^{-\xi} \right] +  \left[\partial_\xi - \Pe_\kappa^{-1} (\partial_\xi^2 - \Hu^2) \right]\left[\Ho +  \frac{Ge^{-\xi}}{1+G}\right]\right\}f_h,
\label{qh}
\end{equation}
and, substituting into \eqref{cont5}, obtain a fourth-order equation for the $\xi$ dependence of the aperture field,
\begin{equation}
(\partial_\xi^2 - \Hu^2)e^\xi \left\{\left[\Da_\kappa \Ho + \Pe_\kappa^{-1}\partial_\xi e^{-\xi} \right] +  \left[\partial_\xi - \Pe_\kappa^{-1} (\partial_\xi^2 - \Hu^2) \right]\left[\Ho +  \frac{Ge^{-\xi}}{1+G}\right]\right\}f_h + 3\Hu^2f_h = 0.
\label{eq:dispersion}
\end{equation}

The boundary conditions on the perturbations can be found from Eqs.~\eqref{eq:cbc} and \eqref{eq:qbc}. From the inlet and outlet conditions \eqref{eq:cbc} it follows that dissolution at the inlet is uniform (because $\Hc=1$), 
\begin{equation}
f_h(\xi=0) = 0,
\label{eq:bc1}
\end{equation}
and that far downstream the aperture is unperturbed,
\begin{equation}
f_h(\xi \rightarrow \infty) = 0.
\label{eq:bc2}
\end{equation}
The boundary conditions on the flow \eqref{eq:qbc} also impose conditions on $f_h$ through Eq.~\eqref{qh}. The uniform pressure at the inlet leads to a condition on $q_\xi$,
\begin{equation}
f_q(\xi=0) = \left[\partial_\xi f_q\right]_{\xi=0}= 0,
\end{equation}
which, by means of \eqref{qh}, imposes a third-order boundary condition on $f_h$,
\begin{equation}
\left[\partial_\xi e^\xi \left\{\left[\Da_\kappa \Ho + \Pe_\kappa^{-1}\partial_\xi e^{-\xi} \right] +  \left[\partial_\xi - \Pe_\kappa^{-1} (\partial_\xi^2 - \Hu^2) \right]\left[\Ho +  \frac{Ge^{-\xi}}{1+G}\right]\right\}f_h \right]_{\xi=0} = 0,
\label{eq:bc3}
\end{equation}
The outlet condition
\begin{equation}
f_q(\xi\rightarrow\infty) = 0,
\label{flow_inf}
\end{equation}
imposes a further restriction on $f_h$, through Eq.~\eqref{qh}, namely that it must decay at least as fast as $e^{-\xi}$,
\begin{equation}
e^{\xi} f_h(\xi\rightarrow\infty) = A.
\label{eq:bc4}
\end{equation}
In most cases the constant $A$ must be zero in order for \eqref{flow_inf} to be satisfied, but in the convective limit ($H = 0$), the solution $f_h = Ae^{-\xi}$ is an eigensolution of \eqref{qh} with zero eigenvalue, and therefore satisfies the far-field boundary condition on $f_q$.

Since the initial amplitude of the instability is arbitrary, the four boundary conditions impose an additional constraint which can be used to solve for the eigenvalue $\Ho(\Hu)$. We have used a spectral method, which we summarize in Sec.~\ref{sec:spectral}, to find the dispersion relation numerically. In certain limiting cases further analysis is feasible; we describe these on a case by case basis in Sec.~\ref{sec:results}

\section{Spectral method}\label{sec:spectral}

The solution of equation \eqref{eq:dispersion},  together with the boundary conditions \eqref{eq:bc1}, \eqref{eq:bc3}, and \eqref{eq:bc4}, was obtained using the pseudospectral, boundary-bordering method~\citep{Boyd1987,Boyd2001}. For a given linear operator, ${\cal H}$, the differential equation 
\begin{equation}
{\cal H} f(\xi) = g(\xi), \ \ \ \ \ \ \ 0 \leq \xi \leq \infty,
\end{equation}
is represented as a linear system
\begin{equation}
{\VH} {\Vf} = {\Vg}
\end{equation}
where the elements of the vector ${\Vf}$ are the coefficients of the expansion of $f(\xi)$ in the basis functions $\Psi_j(\xi)$,
\begin{equation}
f(\xi) = \sum_{j=1}^N f_j \Psi_{j-1}(\xi). 
\end{equation}
Matrix elements of $\cal H$ are calculated at $N-2$ collocation points, $\xi_i$,
\begin{equation}\label{eq:matrix}
H_{i+2,j} = [{\cal H} \Psi_{j-1}(\xi)]_{\xi=\xi_i}
\end{equation}
and the corresponding elements of the right-hand-side vector are
\begin{equation}
g_{i+2} = g(\xi_i).
\end{equation}

The first two rows of $\VH$ are used impose the boundary conditions
at $\xi=0$. If the boundary conditions are expressed in terms of the linear operators ${\cal B}_{i^\prime}$,
\begin{equation}
{\cal{B}}_{i^\prime}(f) = \alpha_{i^\prime}, \ \ \ \ \ \ \ \ i^\prime=1,2,
\end{equation}
then in the matrix representation
\begin{equation}\label{eq:bcCh}
H_{i^\prime,j} = \left[{\cal{B}}_{i^\prime}\Psi_{j-1}(\xi)\right]_{\xi=0}, \ \ \ \ \ \ \ \ g_{i^\prime}=\alpha_{i^\prime},
\end{equation}
where $i^\prime = 1,2$.

The basis functions are rational Chebyshev functions in ${\cal R}^+ = [0,\infty]$, defined as
\begin{equation}
\Psi_n(\xi) = T_n\left(\frac{\xi-L}{\xi+L}\right),
\end{equation}
where $T_n(t)$, with $n = 0, 1, 2, \ldots$, is a Chebyshev polynomial of the first kind, defined in the range $-1 \le t < 1$. The convergence of the solution depends on a suitable choice of the mapping parameter, $L$, which varies somewhat with wavelength. For small numbers of basis functions ($N < 20$), we took $L= 1$ at short wavelengths ($\Hu > 1$) and $L=10$ at long wavelengths ($\Hu < 1$). However, for larger numbers of basis functions ($N > 50$), a constant $L=10$ was suitable for the whole range of wavelengths, $0.01 < \Hu < 10$. For a given $L$ and $N$, the $N-2$ collocation points are~\citep{Boyd1987},
\begin{equation}
\xi_i=L \cot^2 \left(\frac{\pi}{4} \frac{2i-1}{(N-2)}\right), \ \ \ \  i=1,\dots,N-2.
\end{equation}

The dispersion relation can be found by solving the linear system of equations represented by \eqref{eq:matrix}--\eqref{eq:bcCh}, with boundary conditions $f(\xi=0) = 0$ \eqref{eq:bc1} and $\partial_\xi f(\xi=0) = 1$, which fixes the amplitude of the perturbation. Then, we iteratively seek the largest value of $\Ho$ for which the boundary condition in \eqref{eq:bc3} is satisfied and hence find the dispersion relation $\Ho(u)$. There is no need to separately impose the far-field regularity conditions, Eqs.~\eqref{eq:bc2} and \eqref{eq:bc4}, since this is automatically incorporated by the basis functions~\citep{Boyd1987}. We have cross-checked the spectral code with analytic solutions in a number of special cases (see Sec.~\ref{sec:results}), and a Maple version of the spectral code is included in the Supplementary Material.

\section{Results}\label{sec:results}

In general, the dispersion relation \eqref{eq:dispersion} must be solved numerically; for example, using the spectral method described in Sec.~\ref{sec:spectral}. However, in the important limiting case of convection-dominated ($H\rightarrow 0$), reaction-limited ($G\rightarrow 0$) dissolution, it is possible to obtain a tractable analytic dispersion relation, as shown in Sec.~\ref{sec:dispclim}. We can also obtain analytic solutions in other limiting cases, but the solutions are too lengthy to be reproduced in print, although we include Maple workbooks as Supplementary Material. Analytic calculations from Maple~\citep{Maple10} and Mathematica~\citep{WolframResearch2008} were crosschecked with each other and with the spectral code (Sec.~\ref{sec:spectral}) in many cases.

\subsection{Convection-dominated dissolution: $H \rightarrow 0$.}\label{sec:dispclim}

In convection-dominated flows ($H\rightarrow 0$), the Damk\"ohler number on the scale of the penetration length $\Da_\kappa = 1$, and the corresponding P\'eclet number $\Pe_\kappa \rightarrow \infty$. The dispersion relation \eqref{eq:dispersion} then simplifies to
\begin{equation}
(\partial_\xi^2 - \Hu^2)e^\xi \left\{\Ho +  \partial_\xi\left[\Ho +  \frac{Ge^{-\xi}}{1+G}\right]\right\}f_h + 3\Hu^2f_h = 0.
\label{eq:dispclim}
\end{equation}
There is an analytic solution of Eq.~\eqref{eq:dispclim} in terms of a linear combination of three generalized hypergeometric functions $z^\alpha(z-1)_3F_2(\{a_1,a_2,a_3\},\{b_1,b_2\}; z)$, where $a_k$ and $b_k$ are complicated algebraic functions of $G$ and $\Hu$, $z = - G\Ho^{-1}\exp(-\xi)/(1+G)$, and $\alpha$ is a simple function of $\Hu$. As the solution is lengthy and not very informative we do not include it here, but a Maple notebook is included as Supplementary Material.

\begin{figure}
\center\includegraphics[width=0.8\textwidth]{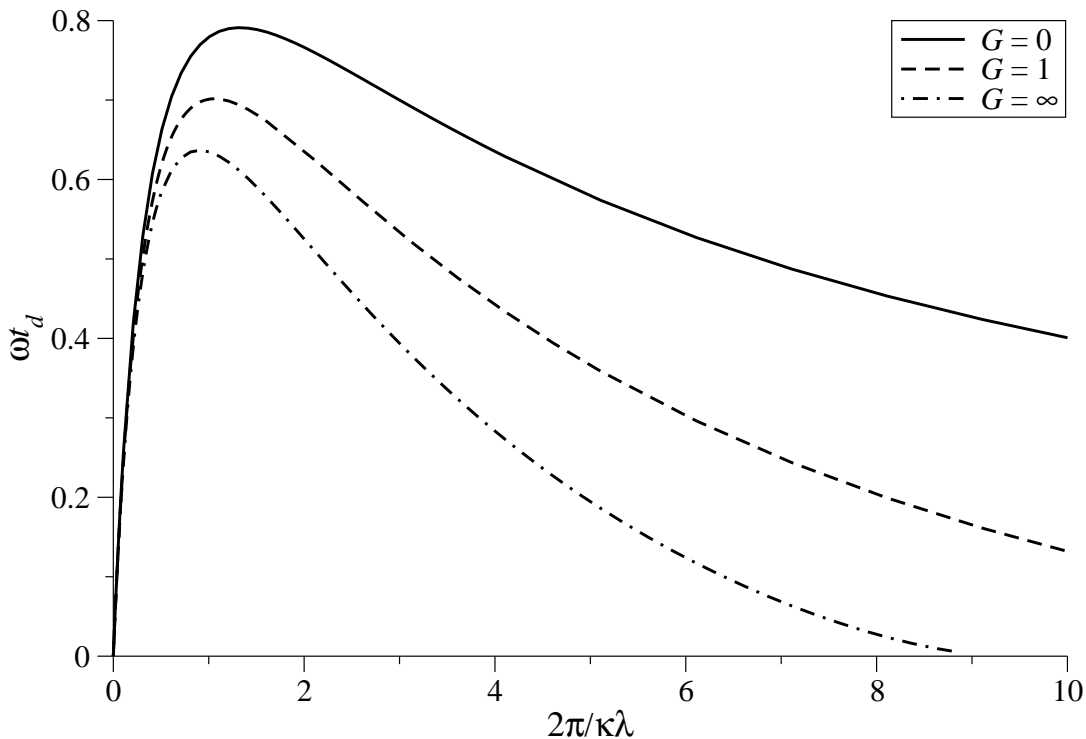}
\caption{Growth rates of the inlet instability in the purely convective case $(H = 0)$. The solid line corresponds to the reaction-limited case ($G = 0$), whereas the dash-dotted curve corresponds to the diffusive limit ($G=\infty$) and the dashed curve is for mixed kinetics ($G=1$). The dimensionless growth rate $\Ho = \omega t_d$, Eq.~\eqref{eq:td}, is plotted against the dimensionless wavevector, $\Hu = 2\pi/\kappa\lambda$.}\label{fig:G}
\end{figure}

A much simpler equation is obtained in the reaction limit ($G \rightarrow 0$) of \eqref{eq:dispclim} \citep{Szymczak2011},
\begin{equation}
(\partial_\xi^2-\Hu^2) \Ho e^\xi (1 + \partial_\xi) f_h + 3 \Hu^2 f_h =0.
\label{eq:climrlim}
\end{equation}
The general solution of \eqref{eq:climrlim} is
\begin{multline}\label{eq:c}
f_h(\xi) = Ae^{-\xi}\, _0F_2\left(1+\Hu, 1-\Hu; 3\Ho^{-1}\Hu^2e^{-\xi}\right) \\ + Be^{(\Hu-1)\xi}\, _0F_2\left(1+\Hu, 1-2\Hu; 3\Ho^{-1}\Hu^2e^{-\xi}\right) \\ + Ce^{-(\Hu+1)\xi}\, _0F_2\left(1+\Hu, 1+2\Hu; 3\Ho^{-1}\Hu^2e^{-\xi}\right),
\end{multline}
where $A$, $B$, and $C$ are constants and $_0F_2(p, q; z)$ is a generalized hypergeometric function. The far field boundary condition \eqref{eq:bc4}  requires that $B = 0$, while the condition $f_h(0) = 0$ \eqref{eq:bc1} is then sufficient to determine the function $f_h(\xi)$ to within an arbitrary constant, which is the initial amplitude of the perturbation. Imposing the final boundary condition \eqref{eq:bc3} gives a dispersion relation for $\Ho(\Hu)$,
\begin{multline}\label{eq:disp}
\left[\Ho^2 \,
_0\tilde{F}_2\left(1+\Hu,1+2\Hu;3\Ho^{-1}\Hu^2\right) +  \right. \\ 3 (1+2\Hu) \Ho \,
   _0\tilde{F}_2\left(2+\Hu,2+2\Hu;3\Ho^{-1}\Hu^2\right)   \\ \left. +  9 \Hu^2 \, _0\tilde{F}_2\left(3+\Hu,3+2\Hu;3\Ho^{-1}\Hu^2\right) \right] {_0\tilde{F}_2}\left(1+\Hu,1-\Hu;3\Ho^{-1}\Hu^2\right) = \\ 3
\left[\Ho \, _0\tilde{F}_2\left(2+\Hu,2-\Hu;3\Ho^{-1}\Hu^2\right) + \right. \\ \left. 3\Hu^2 \, _0\tilde{F}_2\left(3+\Hu,3-\Hu;3\Ho^{-1}\Hu^2\right)\right]\,  _0\tilde{F}_2\left(1+\Hu,1+2\Hu;3\Ho^{-1}\Hu^2\right), 
\end{multline}
where $_0{\tilde F}_2(p,q;z) = {_0F_2}(p,q;z)/\Gamma(p)\Gamma(q)$ is a regularized hypergeometric function \citep{Olver2010}. The maximum growth rate (largest positive root) at each $\Hu$ from \eqref{eq:disp} corresponds to the solid line ($G = 0$) in Fig.~\ref{fig:G}. The positive growth rates show that the front is unstable across the whole spectrum of wavelengths, with a well-defined maximal growth rate, $\Ho_{max} = 0.79 t_d^{-1}$, at a wavelength $\lambda_{max} = 4.74 \kappa^{-1}$.  An individual fracture will therefore develop a strongly heterogeneous permeability during dissolution, with an inherent length scale that depends on the kinetics and flow rate (via $\kappa$), but not the initial topography. There is no lower limit to the reaction rate for unstable dissolution if the scale of the fracture is sufficiently large.

Figure~\ref{fig:G} also shows the impact of reaction kinetics (controlled by the parameter $G$) on the dispersion relation. For wider apertures (\ie $G\gg1$), diffusional transport of reactant across the aperture has a stabilizing effect on the growth of the instability. The fastest-growing wavelength, $\lambda_{max}$, is pushed towards longer wavelengths and at sufficiently short wavelengths perturbations in the front are stable.

\subsection{Reaction-limited dissolution: $G \rightarrow 0$.}\label{sec:disprlim}

The dispersion relation \eqref{eq:dispersion} can also be solved analytically in the reaction limit, $G = 0$; the solution of the dispersion equation,
\begin{equation}
(\partial_\xi^2 - \Hu^2)e^\xi \left\{\left[\Da_\kappa \Ho + \Pe_\kappa^{-1}\partial_\xi e^{-\xi} \right] +  \Ho \left[\partial_\xi - \Pe_\kappa^{-1} (\partial_\xi^2 - \Hu^2) \right]\right\}f_h + 3\Hu^2f_h = 0,
\label{eq:disprlim}
\end{equation} is again a combination of hypergeometric functions $_3F_3(\{a_1,a_2,a_3\},\{b_1,b_2,b_3\}; z) \exp(g\xi)$, where $a_k$, $b_k$ and $g$ are functions of $\Hu$ and $H$, and $z = -\Ho^{-1}\exp(-\xi)$. Again we have included a Maple notebook in the Supplementary Material.

\begin{figure}
\center\includegraphics[width=0.8\textwidth]{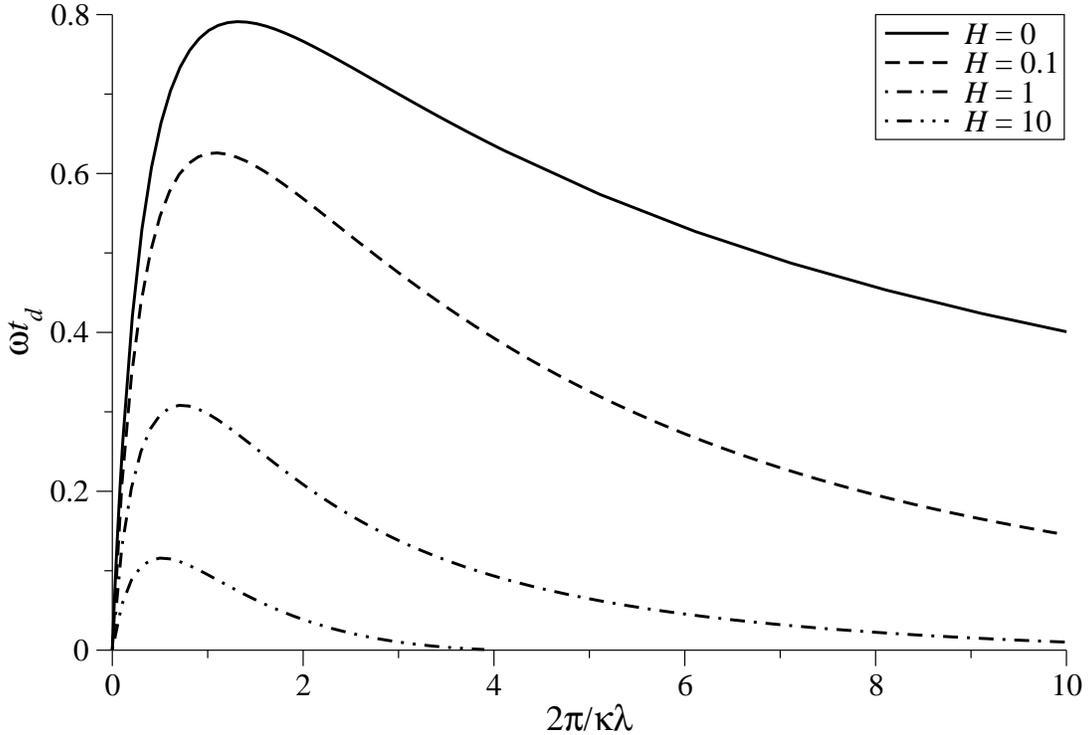}
\caption{Growth rates of the inlet instability in the reaction-limited case $(G=0)$. The solid line corresponds to the purely convective case ($H = 0$). The other lines show results for increasing $H$: $H = 0.1$, $H = 1$, and $H = 10$.  The dimensionless growth rate $\omega t_d$ is plotted against the dimensionless wavevector $2\pi/\kappa\lambda$.}\label{fig:H}
\end{figure}

In the diffusive limit ($H \rightarrow \infty$), $\Da_\kappa \rightarrow \Pe_\kappa^{-1} \rightarrow \sqrt{H}$ and the dispersion relation contains only a single length scale ($\kappa^{-1}$);
\begin{equation}
(\partial_\xi^2 - \Hu^2)e^\xi \left\{\partial_\xi e^{-\xi} - \Ho(\partial_\xi^2 - \Hu^2 -1)\right\}f_h = 0.
\label{eq:disprlim2}
\end{equation}
It is possible to show analytically that the only root of the dispersion relation is $\hat{\omega}=0$, which means that dissolution is neutrally stable in the diffusive limit ($H \rightarrow \infty$). On the other hand, the numerical results in Fig.~\ref{fig:H} imply that the dissolution is unstable for even an infinitesimal convective flux.

\subsection{Geophysical implications}\label{sec:geo}

\begin{figure}
\center\includegraphics[width=1.0\textwidth]{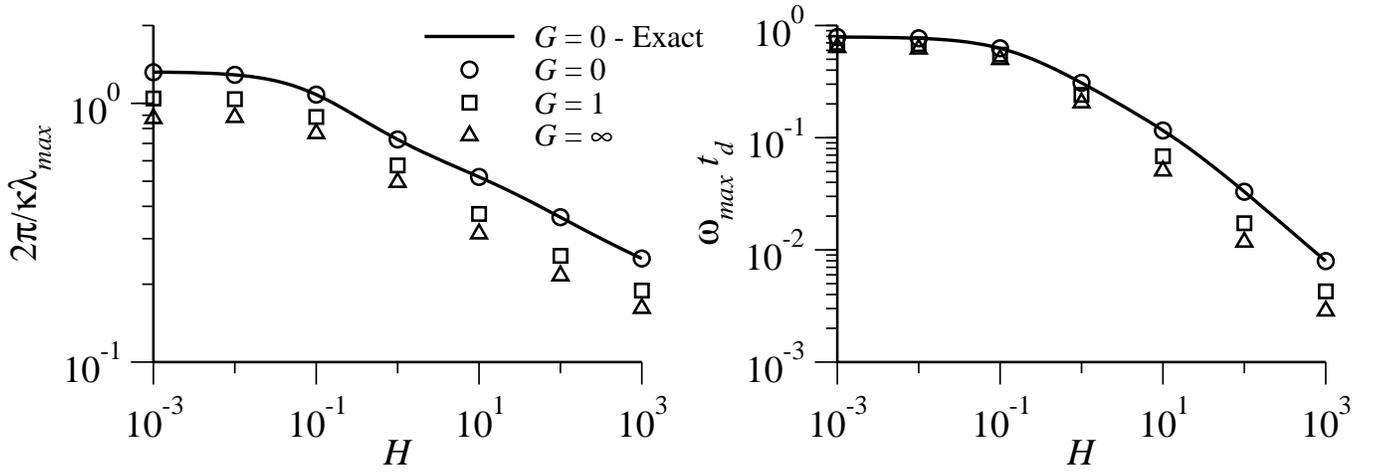}
\caption{Wavelength ($\lambda_{max}$) and growth rate ($\omega_{max}$) of the most unstable mode. The solid lines correspond to the exact solution of the reaction limit \eqref{eq:disprlim}. The symbols indicate numerical results from the spectral code for a range of $G$: $G = 0$ (circles), $G = 1$ (squares), and $G = \infty$ (triangles).  The dimensionless wavevector, $\Hu = 2\pi/\kappa\lambda$, and growth rate, $\Ho = \omega t_d$, are plotted against the convective parameter $H$. The results for $G=0$ agree to at least 6 figures with the analytic dispersion relation from \eqref{eq:disprlim}. Up to 320 basis functions were needed to obtain convergent solutions (6 figures) from the spectral code at large $H$, but as few as 20 are sufficient for $H < 1$.}\label{fig:max}
\end{figure}

Figure~\ref{fig:max} summarizes the most important results of this study. Here we plot the (dimensionless) wavevector and growth rate of the dominant (most unstable) mode of the fracture instability. The convective limit extends up to $H \approx 0.01$; in this range both the dimensionless wavelength and growth rate are nearly constant. Thus for convection-dominated infiltration, the wavelength and timescale are simply related to the underlying geophysical parameters:
\begin{equation}
\lambda_{max} \approx \frac{2.4 q_0}{k_{eff}}, \ \ \ \ \omega_{max} \approx \frac{1.6 k_{eff} \gamma}{h_0}. \label{eq:maxwithdim}
\end{equation}

In order to put these results in a geological context, we consider typical values of the physical parameters characterizing dissolving fractures. Fracture apertures are between $0.005\cm$ and $0.1\cm$ \citep{Motyka1984,Paillet1987,Dreybrodt1996}, and hydraulic gradients are of the order of $10^{-3}$ to $10^{-1}$ \citep{Palmer1991,Dijk1998}. This gives a range of characteristic flow velocities in undissolved fractures from $10^{-4} \cm\s^{-1}$ to $10 \cm\s^{-1}$. The corresponding P\'eclet numbers are $10^{-1} < \Pe < 10^5$, taking the solute diffusion coefficient $D=10^{-5} \cm^2\s^{-1}$. The reaction rates vary widely, depending on the mineral. For example, relatively fast dissolving gypsum has a reaction rate $k$ of the order of $0.01 \cm\s^{-1}$ \citep{Jeschke2001}, whereas siliceous  minerals have surface reaction rates of the order of $10^{-9} \cm\s^{-1}$ \citep{Rimstid1980,Dijk1998}. The typical reaction rates for calcite are in the range $10^{-5} \cm\s^{-1} - 10^{-4} \cm\s^{-1}$  \citep{Palmer1991,Dreybrodt1996}. Thus the limitations imposed by the diffusion of reactant across the fracture aperture vary widely, resulting in a broad range of possible $G$ values: from $G \sim 10^{-7}$ in quartz, through $G \approx 0.1$ for a typical calcite fracture, up to $G \sim 1-10$ in gypsum. Nevertheless, Fig.~\ref{fig:max} shows that both the maximal growth rate and the position of the maximally unstable wavelength depend only weakly on G; $\Ho_{max}$ changes by 25\% in the range $0 < G < \infty$, with a similar change in the corresponding wavelength. However, the data in Fig.~\ref{fig:max} is given in terms of dimensionless quantities and the absolute growth rates vary dramatically across different minerals. For quartz, with $\gamma=6 \cdot 10^{-5}$~\citep{Dijk1998}, the time unit $t_d \sim 5000$ years, whereas the relevant timescale for calcite is a few months~\citep{Szymczak2011a}. The same holds for the instability wavelengths, $\lambda$, which vary from centimeters (gypsum) to kilometers (quartz). It is important to realize that the initial instability wavelength will in general be different from the spacing between protrusions in a mature formation. This is due to a coarsening of the pattern that is characteristic of this kind of dynamics~\citep{Szymczak2006}; the  fingers compete with each other for the flow such that the longer ones grow more rapidly but the shorter ones become stagnant. As a result, the characteristic length between active (growing) protrusions increases with time.

In geophysical systems, diffusion has only a small effect on the instability. Although $H$ can vary from $\sim 10^{-15}$ (for wide fractures in siliceous formations) up to about $1$ for narrow fractures in gypsum, fracture dissolution is typically convection dominated ($H \ll 1$). The residual diffusion leads to a slight shift of the peak growth rate towards longer wavelength, as observed in Fig.~\ref{fig:max}, but the wavelength and growth rate depend primarily on $\Da_{eff}$ \eqref{h0}, via the penetration length $l_p$ and the dissolution time scale $t_d$, with just small corrections from $H$.

These considerations refer to fracture dissolution in a natural geological setting. For carbonate acidization (e.g. with hydrochloric acid) the corresponding reaction rates are significantly higher than for dissolution with aqueous ${\rm CO}_2$; in acidization $k \sim 10^{-1} \cm\s^{-1}$ \citep{Fredd1998}, so that $G$ can be larger than 100 (for $h_0 \approx 0.1 \cm$), which means that the dissolution rate is strongly limited by diffusion across the aperture. In the transport limit ($G \gg 1$), $H = \Sh/\Pe^2$ is small under the typical flow rates used in acidization.

\subsection{Reaction order}\label{sec:ro}

Experiments on the dissolution of limestone suggest that, near saturation, dissolution follows a nonlinear rate law, \cf Eq.~\eqref{eq:Rcunder}:
\begin{equation}
\label{eq:R}
R(c_{ca}) = - k c_{sat} \left(1-\frac{c_{ca}}{c_{sat}}\right)^n, \ \ \ \ \ n>1, 
\end{equation}
where $c_{sat}$ is the saturation concentration of calcium ions. If we define a relative undersaturation $\Hc = (c_{sat}-c_{ca})/(c_{sat}-c_{in})$, where $c_{in}$ is the concentration of calcium ions at the inlet, then the transport equation, from \eqref{eq:CD}, is
\begin{equation}\label{eq:CDnonlin}
q_x \partial_x \Hc + q_y \partial_y \Hc = - 2 k\left(1-\frac{c_{in}}{c_{sat}}\right)^{n-1} \Hc^n.
\end{equation}
For simplicity, we only consider reaction-limited, convection-dominated dissolution. The equation describing aperture opening, analogous to \eqref{eq:erosion}, is
\begin{equation}\label{eq:erosionnonlin}
\partial_t h =  2k \gamma \left(1-\frac{c_{in}}{c_{sat}}\right)^{n-1}  \Hc^n,
\end{equation}
where $\gamma = (c_{sat}-c_{in})/\nu c_{sol}$. The remaining equations, continuity and compatibility, are given by Eqs.~\eqref{eq:continuity} and \eqref{eq:compatibility}.

Assuming the aperture in the base state is uniform, $h_b(x) = h_0$, the base concentration profile is
\begin{equation}
\Hc_b(x) = \left(1+\frac{2 k (n-1) (1-c_{in}/c_{sat})^{n-1} x}{q_0}\right)^{\tfrac{1}{1-n}} = \left[1+ (1-n^{-1}) \kappa x\right]^{\tfrac{1}{1-n}}
\label{pfl}
\end{equation}
where $\kappa = 2kn(1-c_{in}/c_{sat})^{n-1}/q_0$. In the limit $n \rightarrow 1$, Eq.~\eqref{pfl} approaches the exponential base profile for linear reaction kinetics \eqref{expo} and the expression for $\kappa$ reduces to Eq.~\eqref{h0}. 

A dispersion equation for the growth rate can be obtained for non-linear kinetics by following the procedure in Sec.~\ref{sec:lsa}, starting with the analogues of Eqs.~\eqref{transport3}--\eqref{erosion3}:
\begin{equation}
\delta q_x \partial_x \Hc_b + q_b \partial_x \delta \Hc = - \frac{1}{\gamma} \partial_t \delta h,
\label{transport6}
\end{equation}
\begin{equation}
\partial_t \delta h = 2 k \gamma \left(1-\frac{c_{in}}{c_{sat}}\right)^{n-1} n \Hc_b^{n-1} \delta \Hc.
\label{erosion6}
\end{equation}
The continuity and compatibility relations are the same as Eqs.~\eqref{cont3}--\eqref{compat3}.
Introducing dimensionless variables:
\begin{equation}
\xi = \kappa x, \ \ \ \ \ \eta = \kappa y, \ \ \ \ \ \tau = \frac{2k\gamma t (1-c_{in}/c_{sat})^{n-1} }{h_0},
\label{dims1}
\end{equation}
and scaling $\delta h$ and $\Vq$ as in \eqref{dims2}, 
we obtain the following equations for $f_c$, $f_h$, and $f_q$, defined in Eqs.~\eqref{eq:dc}--\eqref{eq:dq}:
\begin{align}
& f_q \partial_\xi \hat{c}_b +  \partial_\xi f_c = - \frac{\Ho}{n} f_h, \label{eq:tnonlin}\\
& \frac{\Ho}{n} f_h =  \hat{c}_b^{n-1} f_c, \label{eq:enonlin}\\
& (\partial_\xi^2 - \Hu^2)f_q = -3\Hu^2f_h.\label{eq:compat}
\end{align}
The inlet saturation, $c_{in}$, has been absorbed into the length and time scales \eqref{dims1}.

The base concentration ($\Hc_b$) can be eliminated from the equations for transport \eqref{eq:tnonlin} and erosion \eqref{eq:enonlin} by using \eqref{pfl}: 
\begin{align}
& f_q = \left[1+ (1-n^{-1}) \xi\right]^{\tfrac{n}{n-1}} \left( \Ho f_h + n \partial_\xi f_c \right)    \\
& \Ho f_h = n\left[1+(1-n^{-1}) \xi\right]^{-1} f_c.
\end{align}
Combining these equations with \eqref{eq:compat} we get a dispersion equation for arbitrary kinetic order,
\begin{equation}
\Ho (\partial_\xi^2-\Hu^2) \left[1+ (1-n^{-1}) \xi\right]^{\tfrac{n}{n-1}} \left[1 + \partial_\xi (1+(1-n^{-1})\xi)\right] f_h + 3 \Hu^2 f_h =0, 
\end{equation}
which is well behaved in the limits $n \rightarrow 1$ and $n \rightarrow \infty$.

\begin{figure}
\center\includegraphics[width=0.8\textwidth]{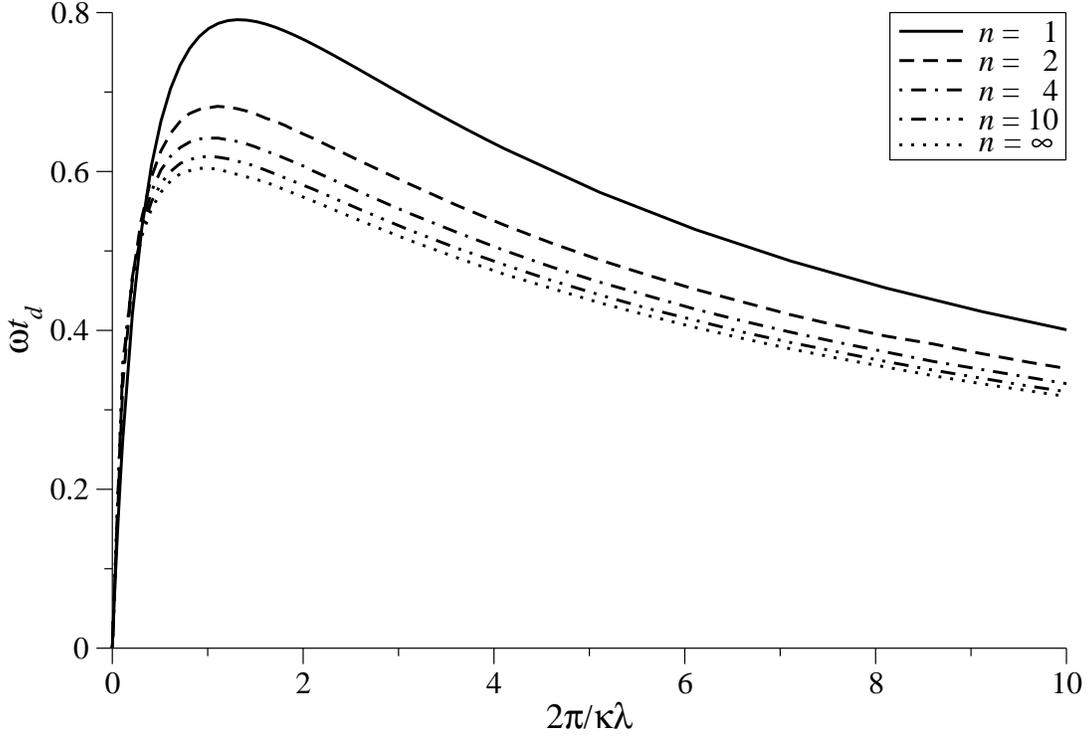}
\caption{The impact of kinetic order on the growth rates. The growth rate of the instability is shown for various powers of $n$, including the linear rate law, $n=1$, and the limit of high reaction order, $n \rightarrow \infty$.}\label{fig:N}
\end{figure}

The impact of kinetic order is illustrated in Fig.~\ref{fig:N}, which shows that even strongly non-linear reaction kinetics ($n \rightarrow \infty$) do not suppress the instability. The dimensionless growth rate depends only weakly on reaction order, reflecting our choice of scaling for the dimensionless length and time. Thus, as a first approximation we can take the peak growth rate as $\Ho_{max} \sim 1$ and the corresponding wavevector $\Hu_{max} \sim 1$, independent of reaction order. Then, in absolute terms, the wavelength corresponding to maximum growth is roughly proportional to $n^{-1}$; $\lambda_{max}^{(n)} \approx \lambda_{max}/n (1-c_{in}/c_{sat})^{n-1}$, where $\lambda_{max} \sim 2\pi q_0/2k$ is the peak wavelength for linear kinetics. This is slightly counterintuitive since increasing reaction order tends to increase the penetration of reactant into the fracture. Nevertheless its effect on the instability is to shorten the wavelength of the most unstable mode. However the wavelength is also strongly dependent on $c_{in}$, and a partially saturated solution at the inlet increases the wavelength of the most unstable mode. The inlet solution to the fracture must be nearly saturated ($c_{in} \rightarrow c_{sat}$) for non-linear kinetics to apply \citep{Plummer1978}, so the wavelength in such cases is almost entirely dependent on the extent of the (small) undersaturation. The corresponding growth rate of the instability $\omega_{max}^{(n)} = \omega_{max} (1-c_{in}/c_{sat})^{n-1}$ is sharply limited by the degree of undersaturation.

\subsection{Finite length fractures}\label{sec:L}

\begin{figure}
\center\includegraphics[width=0.8\textwidth]{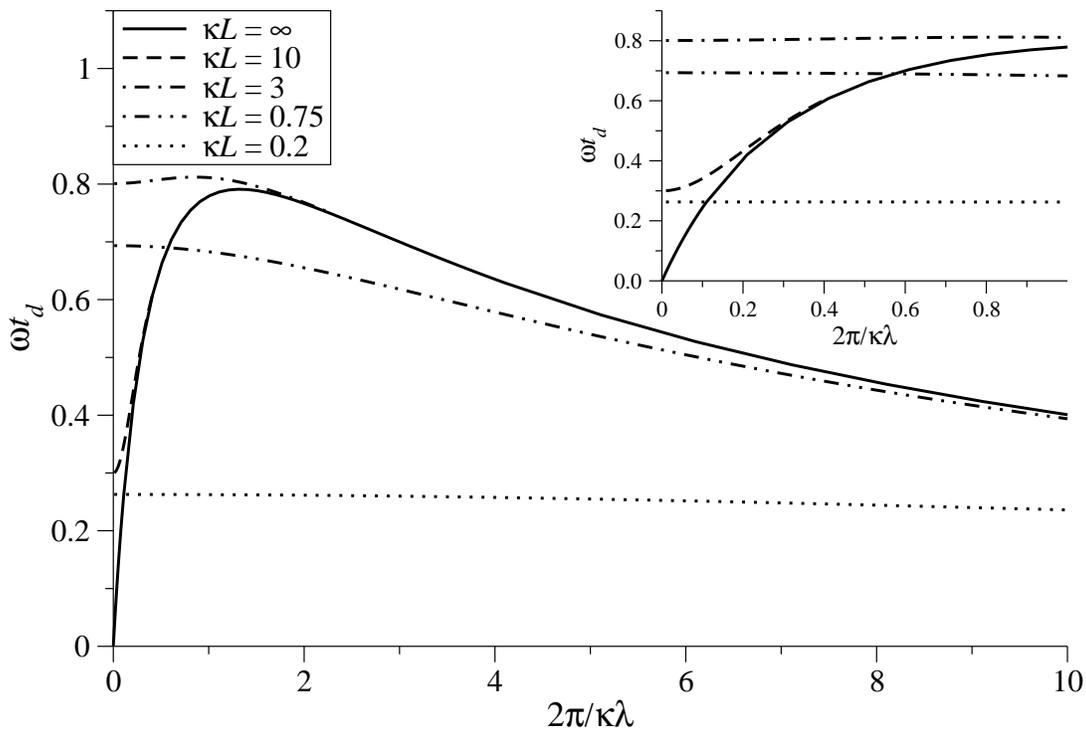}
\caption{The effect of finite system size on the growth rate in the convection-reaction $(G = H = 0)$ limit.}\label{fig:L1}
\end{figure}

The previous analysis corresponds to a semi-infinite system, $x \geq 0$, which is the relevant limit for geophysical systems where the length of the system, $L$, is usually many orders of magnitude larger than the penetration length $\kappa^{-1}$. However, in laboratory experiments as well as in petroleum reservoir stimulation, the relevant length scales are much smaller and finite-size effects may be important. In this case, the far-field boundary condition $q_x(x\rightarrow \infty,y,t)=q_0$ must be replaced by a constant pressure condition at the outlet; then $q_y(x=L,y,t)=0$ or, in terms of perturbations,
\begin{equation}\label{eq:bc3a}
\delta q_y (x=L) = 0.
\end{equation}

\begin{figure}
\center\includegraphics[width=0.8\textwidth]{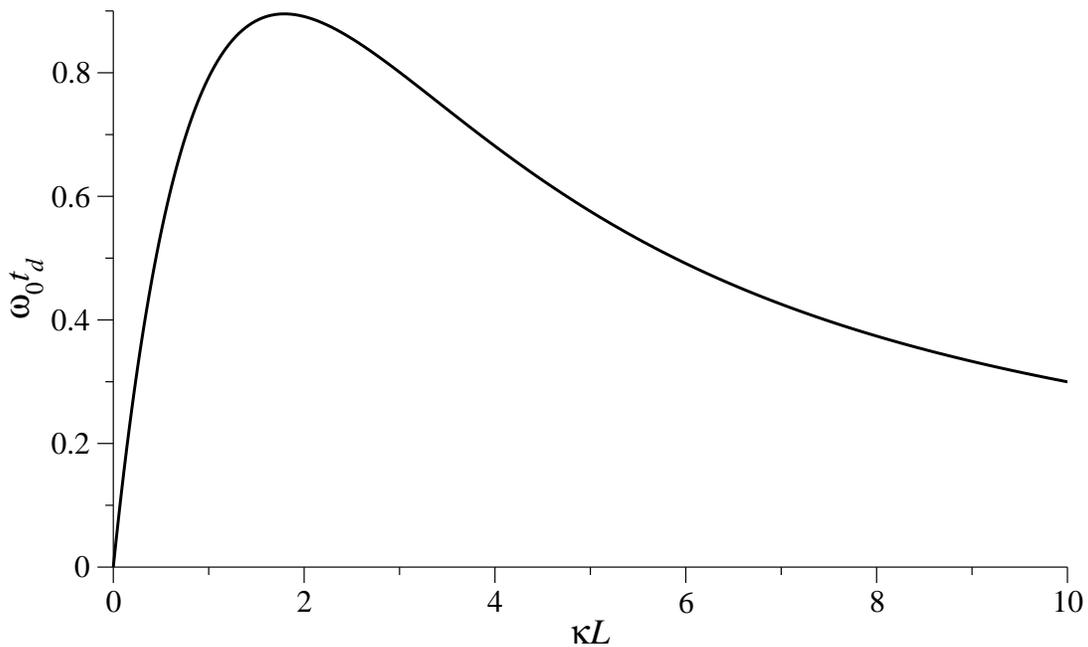}
\caption{Growth rate in the long-wavelength ($\Hu \rightarrow 0$) limit, $\omega_0$, in the convection-reaction $(G = H = 0)$ limit.}\label{fig:L2}
\end{figure}

Figure \ref{fig:L1} shows the effect of a finite length aperture in reaction limited, convection-dominated dissolution ($H=G=0$). Now all three solutions from Eq.~\eqref{eq:c} are needed; Eqs.~\eqref{eq:bc1} and \eqref{eq:bc3a} fix the perturbation to within an arbitrary amplitude, while Eq.~\eqref{eq:bc3} enforces the eigenvalue condition. The additional length scale leads to a richer spectrum of possibilities; in particular, the longest wavelengths are now less stable than in unbounded ($L \rightarrow \infty$) fractures. The shape of the dispersion curve changes considerably as the length of the system is reduced and for short fractures, ($\kappa L < 2$), the growth rate is maximum at zero wavevector. As the length of the fracture increases, the wavelength of the most unstable mode shifts to larger $\Hu$ and the longest wavelengths are only weakly unstable; as $L \rightarrow \infty$ the growth rate at zero wavevector vanishes altogether. In fact, the growth rate at $\Hu=0$ has a particularly simple analytical form
\begin{equation}
\Ho(u=0) = \frac{3 \left[1-(1+\kappa L)e^{-\kappa L}\right]}{\kappa L},
\end{equation}
which is shown in Fig.~\ref{fig:L2}. Both for very small and very large lengths the long-wavelength growth rate is relatively small, with a maximum at $\kappa L \approx 1.8$.

An analysis of Fig.~\ref{fig:L1}, together with Fig.~\ref{fig:G}, offers some insight into the typical dispersion curve for a fracture dissolution instability, which exhibits a strong wavelength selection with a well-defined maximum in the growth rate for $\lambda_{max} \approx \kappa^{-1}$. The results presented in this section show that stabilization of the growth of long wavelength instabilities is connected with the far-field boundary condition \eqref{flow_inf}, which imposes a uniform flow at large distances from the inlet. However, in a finite system, the constant pressure condition at $x=L$ does not require $q_x$ to be uniform, and hence does not lead to a stabilization of long-wavelength modes, as shown in Figs.~\ref{fig:L1} and~\ref{fig:L2}. On the other hand, Fig.~\ref{fig:G} shows that the shape of the short-wavelength spectrum is controlled by reaction kinetics. In particular, transport-limited kinetics decreases the short-wavelength growth rates, since in this regime dissolution slows down as the fracture opens \eqref{eq:keff}.

\section{Conclusions}

In this paper, we have analyzed the stability of a one-dimensional reaction front in dissolving fractures. Strikingly, the dissolution front turns out to be unstable over a wide range of wavelengths, suggesting that fracture dissolution is an inherently two-dimensional process. The maximal growth rate corresponds to wavelengths of the order of the penetration length $\kappa^{-1}$ and this result turns out to be remarkably insensitive to the details of the reaction and transport mechanisms in the fracture: the maximum is shifted towards longer wavelengths when strong diffusion is present or for strongly nonlinear reaction kinetics, but the shift is relatively small and $\kappa \lambda_{max}$ remains within the same order of magnitude. The only case where there is a qualitative change in the dispersion curve is a finite-length system. For relatively short fractures, $\kappa L \leq 3$, the maximum growth rate  occurs at zero wavevector and long-wavelength modes remain unstable.

In summary, the reactive front instability has been shown to be a generic phenomenon in the dissolution of fractured rock. Hence the predictions of fracture breakthrough times, crucial for speleogenesis and for the assessment of subsidence hazards, cannot be based on one-dimensional models. Instead, a two-dimensional model is necessary to take into account the highly localized dissolution front. Numerical ~\citep{Hanna1998,Detwiler2007,Szymczak2011} and theoretical~\citep{Szymczak2006} work has suggested that the dissolutional instability leads to a strong focusing of the fluid flow into a few active channels, which advance in the fracture while competing with each other for the available reactant. However, a quantitative characterization of this non-linear process, which is essential for the prediction of fracture breakthrough times, remains elusive.

\begin{acknowledgments}
This work was supported by the US Department of Energy, Chemical Sciences, Geosciences and Biosciences Division, Office of Basic Energy Sciences (DE-FG02-98ER14853). Computations in this paper were performed by using Maple${13}^\mathrm{TM}$ and Mathematica${7.0}^\mathrm{TM}$.
\end{acknowledgments}

\appendix
\section{Convection-diffusion equation for reactant transport}\label{sec:CD}

In this Appendix we will investigate the validity of the two-dimensional steady-state transport equation \eqref{eq:CD} in various parameter ranges. In the spirit of the Reynolds approximation, we will then assume that the global solution for parallel plates can be applied locally, if the fracture aperture field varies sufficiently slowly, $\left| \nabla h \right| \sim {\cal O}(1)$. Taking the flow to be along the $x$-axis and the normal to the fracture surfaces along the $z$-axis, the convection-diffusion equation for the three-dimensional concentration field $c_{3d}(x,z,t)$ can be written as
\begin{equation}\label{eq:CD3D}
\Dt c_{3d} + v_x \Dx c_{3d} = D \left( \Dx^2 c_{3d} + \Dz^2 c_{3d} \right),
\end{equation}
where $v_x = 6 v_a (z/h - {z^2}/{h^2})$ and $v_a = q_x/h$ is the aperture-averaged fluid velocity. In addition we have boundary conditions on the fracture surfaces
\begin{equation}\label{eq:BC}
D \Dz c_{3d}|_{z=0} = kc_{3d}, ~ D \Dz c_{3d}|_{z=h} = -kc_{3d},
\end{equation}
and at the inlet,
\begin{equation}\label{eq:BCin}
c_{3d}|_{x=0} = c_{in},
\end{equation}
where $c_{in}$ is the inlet concentration.

A direct integration of \eqref{eq:CD3D} over the $z$ coordinate gives a two-dimensional averaged convection-diffusion-reaction equation involving three different concentrations,
\begin{equation}\label{eq:CD2D}
\Dt c_a + v_a \Dx c = D \partial_x^2 c_a -\frac{2k}{h} c_w;
\end{equation}
$c_a(x,t) = h^{-1}\int_0^h c_{3d}(x,z,t) dz$, is the aperture-averaged concentration, $c$ is the cup-mixing concentration \eqref{eq:CD}, and $c_w(x,t) = c_{3d}(x,0,t) = c_{3d}(x,h,t)$ is the reactant concentration at the fracture surfaces. Following standard procedures for averaging the convection-diffusion equation, we will solve Eqs.~\eqref{eq:CD3D}--\eqref{eq:BCin}  to find relations between these average concentration fields in different parameter ranges. In particular we will show that Eq.~\eqref{eq:CD} is correct in the important limits of convection-dominated transport (Sec. \ref{sec:conv}) and reaction-limited (Sec. \ref{sec:rlim}) kinetics.

\subsection{Scaling and steady state}\label{sec:SS}
The steady state approximation in \eqref{eq:CD} can be justified by the time-scale separation between the transport of reactants and the consequent change in fracture aperture. The dissolution time scale is characterized by $\tilde t_d = h/2k\gamma = t_d/(1+G)$ \eqref{eq:td}, where the acid capacity number $\gamma = c_{in}/\nu c_{sol}$ is usually small, because of the high molar concentration of the solid phase. For example calcite contains roughly 25 moles per liter, whereas even a strong acid is rarely used in more than 1 molar concentrations; in the natural dissolution of calcite by atmospheric $\rm CO_2$, $\gamma \sim 10^{-4}$. To see how a small $\gamma$ leads to the steady-state limit we  scale the time by $\tilde t_d$ in addition to the usual scaling of lengths:
\begin{equation}\label{eq:scali}
\xi = \frac{x}{l}, ~ \zeta = \frac{z}{h}, ~ \tau = \frac{t}{\tilde t_d}.
\end{equation}
The axial distance is scaled by the characteristic length $l = v_a h/2k$, and the transverse distance is scaled by $h$. In addition the fluid velocity is scaled by $v_a$ and the concentration by $c_{in}$:
\begin{equation}\label{eq:scald}
\Hv_\zeta = \frac{v_x}{v_a} = 6 \zeta - 6 \zeta^2, ~ \Hc_{3d} = \frac{c_{3d}}{c_{in}}.
\end{equation}

The scaled convection-diffusion equation,
\begin{equation}\label{eq:CD3Ds}
\gamma \partial_\tau \Hc_{3d} + \Hv_\xi \partial_\xi \Hc_{3d} = \tH \partial_\xi^2 \Hc_{3d} + \tG^{-1}\partial_\zeta^2 \Hc_{3d},
\end{equation}
is then be characterized by $\gamma$ and two new dimensionless groups: $\tG  = 2kh/D$ and $\tH = 2kD/v_a^2 h$. $\tG$ and $\tH$ are related to the corresponding parameters defined in the main body of the paper by $\tG = G\Sh$ \eqref{eq:GG}, and $\tH = H(1+G)$ \eqref{eq:HH}. In this appendix we consider the transverse ($z$) direction explicitly and so the Sherwood number does not appear in the defining equations; the ratio of diffusive and reactive fluxes is then characterized by $\tG$ rather than $G$. Since the reactive flux appears in the boundary conditions rather than the underlying equations, the ratio of diffusive and convective fluxes is more naturally defined by $\tH$ rather than $H$.

The steady-state convection-diffusion equation
\begin{equation}\label{eq:CD3Dss}
\Hv_\xi \partial_\xi \Hc_{3d} = \tH \partial_\xi^2 \Hc_{3d} + \tG^{-1}\partial_\zeta^2 \Hc_{3d}
\end{equation}
is reached in the limit $\gamma \rightarrow 0$, and is valid under most circumstances arising in fracture dissolution. The boundary conditions in the dimensionless variables are
\begin{equation}\label{eq:BCs}
\partial_\zeta \Hc_{3d} = \pm \frac\tG{2} \Hc_{3d}, ~~\zeta= 0,1;
\end{equation}
and the average equation for steady-state reactant transport is
\begin{equation}\label{eq:CDsa}
\partial_\xi \Hc = \tH \partial_\xi^2 \Hc_a -\Hc_w.
\end{equation}

\subsection{Convective limit: $\tH=0$.}\label{sec:conv}

Fracture dissolution is usually characterized by small $\tH$, corresponding to the convective limit $\tH \rightarrow 0$ (Sec.~\ref{sec:geo}). The time-independent convection-diffusion equation is then
\begin{equation}\label{eq:CD3Dssc}
\Hv_\xi \partial_\xi \Hc_{3d} = \tG^{-1} \partial_\zeta^2 \Hc_{3d},
\end{equation}
which can be solved by separation of variables, $\Hc_{3d} = f(\xi)g(\zeta)$~\citep{Gupta2001}. The decay in the axial direction is a sum of exponentials, $\exp(-\lambda_n \xi)$, where $\lambda_n$ are related to the positive eigenvalues of the equation
\begin{equation}\label{eq:eigen}
\partial_\zeta^2 g + 16 r^2(\zeta - \zeta^2)g = 0,
\end{equation}
with $r_n = \sqrt{3\tG\lambda_n/8}$. This equation has a single solution that satisfies the symmetry condition $g(0) = g(1)$,
\begin{equation}\label{eq:g}
g(\zeta) = \,_1F_1\left(\frac{1-r}{4},\frac{1}{2};r(2\zeta-1)^2\right)e^{-2r\zeta(\zeta-1)}.
\end{equation}
Applying the boundary conditions from \eqref{eq:BCs} leads to the eigenvalue equation for $r(\tG)$,
\begin{equation}\label{eq:r}
r\left(r-1\right)\,_1F_1\left(\frac{5-r}{4},\frac{3}{2};r\right) + \left(r-\frac\tG{4}\right)\,_1F_1\left(\frac{1-r}{4},\frac{1}{2};r\right) = 0.
\end{equation}

The average equation for the concentration,
\begin{equation}\label{eq:CD2Dsss}
\partial_\xi \Hc = -\Hc_w,
\end{equation}
implies that for a single mode $\lambda \Hc = \Hc_w$ (the same result follows from integrating Eq.~\eqref{eq:eigen} over $\zeta$). Using the Sherwood number to connect $\Hc_w$ and $\Hc$ \eqref{eq:CW},
\begin{equation}\label{eq:Sh}
\Hc_w = \dfrac{\Hc}{1+\tG/\Sh},
\end{equation}
we can relate the eigenvalue $\lambda = 8r^2/3\tG$ to $\Sh$
\begin{equation}\label{eq:Shdef}
\Sh = \dfrac{\lambda \tG}{1-\lambda}.
\end{equation}
Thus the steady-state convection-reaction equation is simply
\begin{equation}\label{eq:CDavg}
\partial_\xi \Hc = - \dfrac{\Hc}{1+\tG/\Sh},
\end{equation}
where $\Sh(\tG)$ is determined from the smallest root of \eqref{eq:r}. 

For reaction-limited kinetics $r \rightarrow 0$, and the hypergeometric functions in Eq.~\eqref{eq:r} can be expanded around $r = 0$; solving for $\tG$ we obtain a quadratic equation for $\lambda$,
\begin{equation}
\tG = \dfrac{8}{3}r^2 + \dfrac{272}{315}r^4 + {\cal O}(r^6) = \lambda \tG + \dfrac{17}{140}\lambda^2 \tG^2,
\end{equation}
with a solution $\lambda = 1 - 17\tG/140 + {\cal O}(\tG^2)$. The concentration is nearly uniform across the aperture and decays axially as a single exponential $e^{-\lambda\xi}$. From Eq.~\eqref{eq:Shdef} we find the Sherwood number for reaction-limited kinetics $\Sh^0 = 140/17 \approx 8.24$.

In the transport limit the concentration at the walls vanishes (Graetz problem) and the eigenvalues $\lambda_n = 8 r_n^2/3\tG$ can be found from the roots of the equation
\begin{equation}
\,_1F_1\left(\frac{1-r}{4},\frac{1}{2};r\right) = 0.
\end{equation}
The transport-limited Sherwood number, $\Sh^\infty \approx 7.541$, follows from the smallest eigenvalue $r_0 \approx 1.6816$. In the numerical work we will ignore the weak dependence of Sherwood number on $\tG$ and take $\Sh = 8$ throughout.

\subsection{Reaction-limit: $\tG \rightarrow 0$.}\label{sec:rlim}

Away from the convective limit, the diffusive flux prevents a solution of the transport equation \eqref{eq:CD3Dss} by separation of variables. However, when the reaction rate is small, such that $\tG \ll 1$, the deviation in concentration from the average concentration, $c_{3d} - c_a$, can be expanded in powers of $\tG$ \citep{Balakotaiah2004,Balakotaiah2010},
\begin{equation}\label{eq:cG}
\Hc_{3d} - \Hc_a =  \tG c^{(1)} + \tG^2 c^{(2)} + \ldots;
\end{equation}
it follows that
\begin{equation}\label{eq:ci}
\int_0^1 c^{(i)} d\zeta = 0.
\end{equation}
From Eq.~\eqref{eq:CD3Dss}, the zeroth order convection-diffusion equation is
\begin{equation}\label{eq:CD3Dss0}
(6\zeta - 6 \zeta^2) \partial_\xi \Hc_a = \tH \partial_\xi^2 \Hc_a + \partial_\zeta^2 c^{(1)}.
\end{equation}
Integrating Eq.~\eqref{eq:CD3Dss0} across the aperture and using the boundary condition \eqref{eq:BCs} $\partial_\zeta \Hc^{(1)} = \pm c_a/2$, we obtain the average equation 
\begin{equation}\label{eq:CDrlim}
\partial_\xi \Hc_a = \tH \partial_\xi^2 \Hc_a - \Hc_a,
\end{equation}
which is the reaction limit of \eqref{eq:CDsa}. In this limit the concentration profile is uniform across the aperture and all three concentrations, $c$, $c_a$, and $c_w$ are equal.

Equation \eqref{eq:CDrlim} can be subtracted from \eqref{eq:CD3Dss0} to eliminate the diffusion term,
\begin{equation}\label{eq:A26}
(6\zeta - 6 \zeta^2 - 1) \partial_\xi \Hc_a - c_a = \partial_\zeta^2 c^{(1)}.
\end{equation}
Solving for $c^{(1)}$,
\begin{equation}\label{eq:CDc1}
c^{(1)} = \partial_\xi \Hc_a \left(\zeta^3 - \frac{1}{2} \zeta^4 - \frac{1}{2} \zeta^2 + \frac{1}{60} \right) - \Hc_a \left(\frac{\zeta^2}{2} - \frac{\zeta}{2} + \frac{1}{12} \right);
\end{equation}
the linear term in $\zeta$ is introduced to satisfy the boundary conditions in \eqref{eq:BCs} and the constant term is to enforce the condition in \eqref{eq:ci}. Finally, we use Eq.~\eqref{eq:CDc1} to relate $\Hc$ and $\Hc_w$ to $\Hc_a$:
\begin{eqnarray}
\Hc &=& \int_0^1 (6 \zeta - 6 \zeta^2) (\Hc_a + \tG c^{(1)}) d\zeta = \left(1 + \frac\tG{60}\right)\Hc_a - \frac{\tG}{210}\partial_\xi \Hc_a, \label{eq:cm} \\
\Hc_w &=& \Hc_a + \tG c^{(1)}(\zeta=0) = \left(1 - \frac\tG{12}\right)\Hc_a + \frac{\tG}{60}\partial_\xi \Hc_a. \label{eq:cw}
\end{eqnarray}
These are the equivalents of the results in \cite{Balakotaiah2010} (67c \& d), but for flat plates instead of tubes.

Using Eqs.~\eqref{eq:cm} and \eqref{eq:cw} to eliminate $c_a$ and $c_w$ from the average equation \eqref{eq:CDsa}, the transport equation becomes
\begin{equation}\label{eq:A32}
\partial_\xi \Hc = \tH\left(1-\frac{4\tG}{105}\right) \partial_\xi^2 \Hc + \frac{\tH\tG}{210} \partial_\xi^3 \Hc - \left(1-\frac{17 \tG}{140}\right)\Hc.
\end{equation}
The third-order term in Eq.~\eqref{eq:A32},
\begin{equation}\label{eq:A33}
\frac{\tH\tG}{210} \partial_\xi^3 \Hc = \frac{v_a h}{2k}\frac{h^2}{210} \partial_x^3 \Hc,
\end{equation}
is small compared with the convective term,
\begin{equation}\label{eq:A34}
\partial_\xi \Hc = \frac{v_a h}{2k}\partial_x \Hc,
\end{equation}
on all scales larger than the aperture $h$. Since $h$ is small on scales of interest in fracture dissolution we can safely ignore this term. Similarly, the diffusive term $(4\tH\tG/105) \partial_\xi^2 \Hc$ is small compared to $\Hc$. 
Dropping these terms leaves the renomalization of the reaction term as the leading-order correction for finite $\tG$ (in the steady-state limit),
\begin{equation}\label{eq:A35}
\partial_\xi \Hc = \tH \partial_\xi^2 \Hc - \frac{\Hc}{1+\tG/\Sh^0}.
\end{equation}
The average equation for the cup-mixing concentration has no Taylor dispersion term, but only the contribution from molecular diffusion. This is true both in the convective limit (arbitrary $\tG$) and the reaction limit (arbitrary $\tH$). 

\subsection{Summary}
In this appendix we have examined the structure of the depth-averaged convection-diffusion equation across a range of Damk\"ohler and P\'eclet numbers. The dimensionless parameter $H = \Da_{eff}/\Pe$ is usually small in fracture dissolution, which implies a convection-dominated process. In such cases the steady-state convection-reaction equation \eqref{eq:CDavg} follows (see Sec.~\ref{sec:conv}), with only a weak dependence of the Sherwood number on reaction rate and entrance length.

When diffusion plays a significant role, the structure of the average equations is more complex, and it is not possible to rigorously treat transport in the case of significant transverse and axial diffusion ($\tG \gg 1$, $\tH \gg 1$) without considering more than one average concentration~\citep{Balakotaiah2004,Balakotaiah2010}.
Nevertheless, in Sec.~\ref{sec:rlim} we showed that in the reaction limit ($G \ll 1$) the structure of Eq.~\eqref{eq:CDavg} is preserved \eqref{eq:A35}.

\section{Scale-dependent P\'eclet and Damk\"{o}hler numbers}\label{sec:A2}
The one-dimensional transport equation \eqref{eq:CD1D} can be non-dimensionalized by the penetration length $\kappa^{-1}$, 
\begin{equation}\label{eq:CD1Ds}
q_0 \kappa \partial_\xi c - D h_0 \kappa^2 \partial_\xi^2 c = - \frac{2 kc}{1 + G},
\end{equation}
where $\xi = \kappa x$. Dividing Eq.~\eqref{eq:CD1Ds} by $q_0 \kappa$ suggests two new dimensionless constants:
\begin{equation}
\Pe_\kappa = \frac{q_0}{D \kappa h_0} = \frac{\Pe}{\kappa h_0}, ~~ \Da_\kappa = \frac{2k}{q_0 \kappa (1+G)} = \frac{\Da_{eff}}{\kappa h_0}.
\end{equation}
$\Pe_\kappa$ is the ratio of convective to diffusive fluxes on the length scale $\kappa^{-1}$, while $\Da_\kappa$ is the ratio of convective to reactive fluxes on the same scale; $\Da_\kappa$  is based on the effective reaction rate $k_{eff}$ \eqref{eq:keff}. The transport equation on the scale of the penetration length $\kappa^{-1}$ is then
\begin{equation}
\partial_\xi c - \Pe_\kappa^{-1} \partial_\xi^2 c = - \Da_\kappa c.
\end{equation}

The parameter $H$ retains the same meaning with the new definitions of P\'eclet and Damk\"{o}hler number,
\begin{equation}
H = \frac{\Da_{eff}}{\Pe} = \frac{\Da_\kappa}{\Pe_\kappa},
\label{eq:HH2}
\end{equation}
and the two new parameters can be written solely in terms of $H$:
\begin{equation}\label{eq:PekDak}
\Pe_\kappa = \frac{2}{\sqrt{1+4H}-1}, \ \ \ \ \Da_\kappa = \frac{2H}{\sqrt{1+4H}-1}.
\end{equation}
Although $\Pe_\kappa$ and $\Da_\kappa$ are not independent, $\Da_\kappa = 1 + \Pe_\kappa^{-1}$, it is a notational convenience to treat them so; however the results are discussed in terms of the independent parameters $G$ and $H$.

On the relevant length scale for fracture dissolution, $\kappa^{-1}$, the ratio of convective and diffusive fluxes is characterized by $\Pe_\kappa$. Nevertheless we prefer to characterize the dissolution in terms of $G$ and $H$ rather than $G$ and $\Pe_\kappa$, since both $\Pe_\kappa$ and $\Da_\kappa$ have simple expressions in terms of $H$. In the convective limit (the most important for fracture dissolution) $H \rightarrow \Pe_\kappa^{-1}$, while in the diffusive limit $H \rightarrow \Pe_\kappa^{-2}$. Thus the convective limit implies $\Pe_\kappa \rightarrow \infty$ and $H \rightarrow 0$, while the diffusive limit is the opposite, but the mapping is not a simple inverse relation.

\section{Derivation of the compatibility relation}\label{compder}

Throughout the paper we will frequently make use of the compatibility relation \eqref{eq:compatibility}, which can be derived by noting that, from \eqref{eq:Reynolds},
\begin{equation}
\partial_y q_x = - \frac{1}{12 \mu} h^3 \partial_{xy} p - 3 \frac{1}{12 \mu} h^2 \partial_y h \partial_x p =
 - \frac{1}{12 \mu} h^3 \partial_{xy} p + \frac{3}{h} q_x \partial_y h.
\label{k1a}
\end{equation}
Similarly
\begin{equation}
\partial_x q_y = - \frac{1}{12 \mu} h^3 \partial_{xy} p - 3 \frac{1}{12 \mu} h^2 \partial_x h \partial_y p =
 - \frac{1}{12 \mu} h^3 \partial_{xy} p + \frac{3}{h} q_y \partial_x h.
\label{k2a}
\end{equation}
Subtracting \eqref{k2a} from \eqref{k1a} leads to the compatibility relation
\begin{equation}
\partial_y q_x - \frac{3}{h} q_x \partial_y h  = \partial_x q_y - \frac{3}{h} q_y
\partial_x h.
\end{equation}

\end{document}